\documentclass[12pt]{iopart}

\usepackage{graphicx}

\newcommand{\cfig}[2]{\begin{figure}[!hbtp]\centering% !ht 
\includegraphics[width=.4\textwidth]{#1}\caption{\small{#2}}\end{figure}}

\usepackage{iopams}

\begin{document}

\title[The Atomic hypothesis]
{The Atomic hypothesis: Physical consequences}

\author{Mart\'{\i}n Rivas}
\address{Theoretical Physics Department, The University of the Basque Country,\\ 
Apdo.~644, 48080 Bilbao, Spain}
\ead{martin.rivas@ehu.es}

\begin{abstract}
The hypothesis that matter is made of some ultimate and indivisible objects, together
the restricted relativity principle, establishes a constraint on the kind of variables
we are allowed to use for the variational description of elementary particles. We consider that 
the atomic hypothesis not only states the indivisibility of elementary particles, 
but also that these ultimate objects, 
if not annihilated, cannot be modified by any interaction so that all allowed states of an elementary particle
are only kinematical modifications of any one of them. Terefore, an elementary particle 
cannot have excited states.
In this way, the kinematical group of spacetime symmetries not only defines the symmetries of the 
system, but also the variables in terms of which the mathematical description of the 
elementary particles can be expressed in either the classical or the quantum mechanical description. 
When considering the interaction of two Dirac particles, 
the atomic hypothesis restricts the interaction Lagrangian to a kind
of minimal coupling interaction.
\end{abstract}


%\noindent{\it J. Phys. A: Math. Theor.}{\bf 40} (2007) 1-12

\maketitle

%%%%%%%%%%%%%%%%%%%%%%%%%%%%%%%%%%
\section{Introduction}
%%%%%%%%%%%%%%%%%%%%%%%%%%%%%%%%%%

Feynman, in the first chapter of his {\it Lectures on Physics} \cite{Feynman}, states that
{\it "If, in some cataclysm, all of scientific knowledge were to be destroyed, and only one sentence passed 
on to the next generations of creatures, what statement would contain the most information in the fewest words?
 I believe it is the atomic hypothesis (or the atomic fact or whatever you wish to call it)
that all things are made of atoms-little particles that move around in perpetual motion, 
attracting each other when they are a little distance apart, but repelling upon being squeezed into one another."}

If the atomic hypothesis is such an important principle, physics has to take advantage of this 
fact, and, properly formulated, should be included as a preliminary fundamental principle of elementary particle
physics. The aim of this contribution is to show that the atomic hypothesis
not only states the indivisibility of elementary particles, but also that these ultimate objects, 
if not annihilated, cannot be modified so that all states of an elementary particle
are only kinematical modifications of any one of them and, therefore, no excited states are allowed. 

This article is organized as follows. In the next section we give elementary geometrical arguments
leading to the plausible conclusion that the most general motion of the centre-of-charge of a
classical elementary spinning particle is a helical motion at the speed of light, 
so that the location of the charge satisfies,
in general, a fourth-order differential equation, which is the most general 
differential equation satisfied by a 
point in three-dimensional space. 

In the section \ref{sec:part} we analyse three fundamental principles, namely the restricted relativity
principle, the atomic principle and the variational principle, which allow us to obtain a completely
general formalism for describing, at the classical level, elementary spinning particles \cite{Rivasbook}.
The quantization of this formalism is obtained by replacing the variational principle by the uncertainty principle
in the form postulated by Feynman, i.e. in terms of the path integral approach. 

In section \ref{kin} we summarize how the above fundamental principles produce a general kinematical formalism
for describing elementary particles. Dirac's equation is obtained
when quantizing precisely the classical system whose charge is moving along a helix, at the speed of light,
as suggested by the above elementary arguments. 
The main features of a classical Dirac particle are outlined in section 5. 

The importance of the atomic principle is stressed in section \ref{sec:physics}, when analysing the interaction
of two Dirac particles. The atomic principle restricts the dependence of the interaction Lagrangian 
to the positions and velocities of both particles, but not to the accelerations and angular velocities.
It is suggesting a kind of minimal coupling interaction between the currents of both particles.
Finally, section \ref{sec:predict} is devoted to some predictions of the kinematical formalism. 
It is shown that there is a diference
in chirality between matter and antimatter, at the classical level. Matter is left-handed while antimatter
is right-handed. It is also predicted that particles and antiparticles must necesarilly have the same
relative orientation between the spin and magnetic moment. The analysis of a very close electron-electron
interaction shows that, if certain boundary conditions are fulfilled, two electrons with their
spins parallel can form, from a classical point of view, a metastable bound state. We finish with some
final conclusions.

%%%%%%%%%%%%%%%%%%%%%%%%%%%%%%%%%%
\section{Helical motion of the charge of an elementary spinning particle}
%%%%%%%%%%%%%%%%%%%%%%%%%%%%%%%%%%
As is well known in differential geometry, a continuous and differentiable curve in three-dimensional space, 
${\bi r}(s)$, has associated three orthogonal unit vectors, ${\bi t}$,
${\bi n}$ and ${\bi b}$, called respectively the tangent, normal and binormal. 
If using the arc length $s$ as the curve parameter, they satisfy the Frenet-Serret equations
\[
\dot{\bi t}=\kappa{\bi n},\quad \dot{\bi n}=-\kappa{\bi t}+\tau{\bi b},\quad
\dot{\bi b}=-\tau{\bi n},
\] 
where the overdot means $\dot{}\equiv d/ds$. The knowledge of the curvature $\kappa(s)$ and torsion $\tau(s)$, 
together the boundary values ${\bi r}(0)$, ${\bi t}(0)$,
${\bi n}(0)$ and ${\bi b}(0)$, completely
determine the curve, because the above equations are integrable. 
If we call ${\bi r}^{(k)}(s)\equiv d^k{\bi r}/ds^k$, and in particular $\dot{\bi r}\equiv {\bi r}^{(1)}={\bi t}$, 
and eliminate the three unit vectors
among the succesive derivatives ${\bi r}^{(k)}$, $k\ge 1$, one obtains that the most general 
differential equation satisfied by the point ${\bi r}$,
is the fourth order differential system
 \begin{equation}
{\bi r}^{(4)}-\frac{2\dot{\kappa}\tau+\dot{\tau}\kappa}{\kappa\tau}{\bi r}^{(3)}
+\left(\kappa^2+\tau^2+\frac{\dot{\kappa}\dot{\tau}-\tau\ddot{\kappa}}{\kappa\tau}+\frac{2\dot{\kappa}^2
}{\kappa^2}\right){\bi r}^{(2)}
+\frac{\kappa}{\tau}(\dot{\kappa}{\tau}-\dot{\tau}\kappa){\bi r}^{(1)}=0.
 \label{eq:masgeneral}
 \end{equation}
Let us consider that an elementary particle is a localized mechanical system. By localized we mean that,
at least, is described by the evolution of a single point ${\bi r}$. 
This point could be the centre-of-mass, but in order to determine the external forces to obtain
the centre-of-mass evolution, we also need
to know the location of the centre-of-charge to compute the actions of the external fields. 
Let us assume that the elementary particle is charged. 
Its electric structure can be reduced to the location of the centre-of-charge
${\bi r}$ and the subsequent multipoles located at this point. 
If its electric field is spherically symmetric we are reduced
to know the evolution just of the centre-of-charge. 
We do not make the assumption that the centre-of-mass
and the centre-of-charge are necessarily the same point. 
As we shall see this is not true for spinning particles, where the centre-of-mass evolution
is some average of the evolution of the centre-of-charge.

By the previous arguments, the centre-of-charge of an elementary particle will satisfy, in general, 
a fourth order differential equation of the form (\ref{eq:masgeneral}) where $\kappa(s)$ and $\tau(s)$
will depend on the external interaction. 
Let us assume now that the motion of the particle is free. 
This means that we cannot distinguish one point of the evolution from another, 
so that the above equations (\ref{eq:masgeneral}) 
must be explicitely independent of the parameter $s$.
The curvature and torsion are necessarily constants of the motion.
Thus $\dot{\kappa}=\dot{\tau}=0$, and, in the free case, these equations
are reduced to
\[
{\bi r}^{(4)}+\left(\kappa^2+\tau^2\right){\bi r}^{(2)}=\frac{d^2}{ds^2}\left({\bi r}^{(2)}+\left(\kappa^2+\tau^2\right){\bi r}\right)=0.
\] 
If the curvature and torsion are constant the curve is a helix, which can be factorized in terms of a
central point 
\[
{\bi q}={\bi r}+\frac{1}{\kappa^2+\tau^2}{\bi r}^{(2)}, \quad \frac{d^2{\bi q}}{ds^2}=0,
\]
which is moving in a straight
trajectory, while the point ${\bi r}$ satisfies
\[
{\bi r}^{(2)}+(\kappa^2+\tau^2)({\bi r}-{\bi q})=0,
\]
an isotropic harmonic motion of frequency $\omega=\sqrt{\kappa^2+\tau^2}$,
around point ${\bi q}$. 
The point ${\bi q}$ clearly represents the centre-of-mass position of a free particle.
Going further, let us assume that the free evolution is analysed by some inertial observer. Then this 
observer cannot distinguish one instant from another, so that, 
the arc length $ds=|{\bi u}|dt$, where ${\bi u}=d{\bi r}/dt$ is the velocity
of the point, must be also independent of the time $t$. The centre-of-charge of a free elementary
particle is describing a helix at a constant velocity for any inertial observer.

If we make a nonrelativistic analysis, the relationship of the velocity measurements among two arbitrary
inertial observers $O$ and $O'$, is given by ${\bi u}'=R{\bi u}+{\bi v}$, where ${\bi v}$ is the constant 
velocity of $O$ as measured by $O'$ and the constant rotation matrix $R$ is their relative orientation.
Now,
\[
{u'}^2=u^2+v^2+2{\bi v}\cdot R{\bi u}.
\]
If $u'$ has to be also constant for observer $O'$, irrespective of ${\bi v}$ and of the rotation matrix $R$, 
this means that the vector ${\bi u}$ must be a constant vector. The centre-of-charge necessarily moves along
a straight trajectory at a constant velocity, for every inertial observer, and the above general
helix degenerates into a straight line and ${\bi q}={\bi r}$. 
This is the usual description of the spinless or pointlike free
elementary particle, whose centre-of-charge and centre-of-mass are represented by the same point.

However, in a relativistic analysis, there is one alternative not included in the nonrelativistic
approach. The possibility that the charge of an elementary particle will be moving at the speed of light
and, in that case, $u=u'=c$, for any inertial observer. 
This means that the center of the helix is always moving at a velocity $|d{\bi q}/dt|<c$, and, 
as will be shown, it represents the centre-of-mass,  this particle is a massive particle. 
In a variational description of this system
the Lagrangian should depend up to the acceleration of the point ${\bi r}$ 
in order to obtain fourth-order differential equations. This dependence on the acceleration 
will give a contribution to the spin of the particle. The motion of the charge around the centre-of-mass 
produces the magnetic moment of the particle.

In summary, there are only two possibilities for a free motion of the charge of an elementary particle.
One, the charge is moving along a straight line at any constant velocity, and the system has no magnetic moment.
In the other, the particle has spin and magnetic moment, the centre-of-mass and centre-of-charge are different points
and the charge moves along a helix at the speed
of light. Because all known elementary particles, quarks and leptons, are spin $1/2$ particles, we are left
only with the last possibility. This is consistent with Dirac's theory of the electron, 
because the eigenvalues of the components of Dirac's velocity operator are $\pm c$ and we can interpret the
corresponding point as representing the centre-of-charge.

This last possiblity is the description of the centre-of-charge of a relativistic spinning 
elementary particle obtained in the proposed general kinematical formalism \cite{Rivasbook}, 
and which satisfies Dirac's equation when quantized.

In this formalism Dirac particles are localized and also orientable mechanical systems. By orientable
we mean that we have to attach to the above point ${\bi r}$, which represents the position of the charge,
a local cartesian frame to describe its spatial orientation. The rotation of the frame will also contribute
to the total spin of the particle. When quantizing the system, 
the spin $1/2$ is coming from the presence of the orientation variables.
Otherwise, if there are no orientation variables, no spin $1/2$ structure is described 
when quantizing the system.
The dependence of the Lagrangian on the acceleration is necessary for the particle to have magnetic moment
and for the separation between the centre-of-mass and centre-of-charge.
 
%%%%%%%%%%%%%%%%%%%%%%%%%%%%%%%%%%
\section{Fundamental principles}
\label{sec:part}
%%%%%%%%%%%%%%%%%%%%%%%%%%%%%%%%%%
The restricted relativity principle states that, in absence of gravitation, 
there exists a set of equivalent observers, historically called
{\it inertial observers}, for whom the laws of physics must be the same. 
This statement is an empty statement
if not complemented with the assumption that
the way two equivalent observers relate the measurement of any
physical magnitude depends only of how they relate the measurements of spacetime events.
They are thus defined with respect to each other by
a spacetime transformation. 
The set of these transformations for all observers form a group, 
the kinematical group, which must be defined as the
fundamental mathematical object of the formalism.
It is this {\it geometrization} of spacetime which establishes the mathematical framework of the
relativity principle. 

The atomic principle admits that matter cannot be divided indefinitely. 
After a finite number of steps in the division of a portion of matter 
we reach an ultimate object, an {\it elementary particle}. 
In this way all known matter is finally made of these atom-little particles.
Then, what is the difference between an elementary particle
and any other little system? We need to distinguish theoretically a true elementary particle
from a bound system of elementary particles. Otherwise the atomic hypothesis will be also
an empty statement. This requires a proper definition of an elementary particle.
The idea is that an electron, if not annihilated with its antiparticle, always remains an electron
in any process of interaction. It thus means
that an elementary particle has no excited states and, 
if not destroyed, we can never modify its internal structure, so that
all possible states are only kinematical modifications of any one of them. 
If the state of an elementary particle changes, it is always possible to find
another inertial observer who describes the particle in the same state as in the previous instant. 

The variational principle recognizes that the 
the action of the evolution of any mechanical system between some initial 
and final states must be stationary.
This completes the classical framework. For the quantum description we must 
susbtitute this last variational 
principle by the uncertainty principle, in the form proposed by Feynman: all paths of the evolution 
of any mechanical system 
between some initial and final states are equally probable. 
For each path a probability amplitude is defined,
which is a complex number of the same magnitude but whose phase is the action of the system between 
the end points $x_1$ and $x_2$ along the corresponding path.  
Feynman's total probability amplitude $K(x_1,x_2)$ is the sum, or path integral, 
of the probability amplitudes
for all paths joining these points. If we call kinematical variables to these 
classical variables which define the initial
and final states of the variational description, these variables become, after quantization, 
the arguments of the wave function. In this way, classical
and quantum mechanics are described in terms of exactly the same set of classical variables 
and its dynamics in terms of initial and final kinematical states. We want to emphasize the importance
of the identification of the kinematical variables, and the interest of rewriting the Lagrangian formalism
in terms of these variables.

%%%%%%%%%%%%%%%%%%%%%%%%%%%%%%%%%%%%%%%
\section{The kinematical formalism}
\label{kin}
%%%%%%%%%%%%%%%%%%%%%%%%%%%%%%%%%%%%%%%

The definition of elementary particle implies that its states
can be described by a finite set of variables. Let us represent by $x_1$ the values of the
fixed set of variables which define the initial variational state, and, similarly, by $x_2$ 
the final values of these variables. If the system is elementary, 
the final state $x_2$ is a kinematical modification of $x_1$, 
so that there will exist some kinematical group element $g$ such that $x_2=gx_1$, for any $x_1$ and $x_2$. 
The kinematical variables, which define the initial and final states of the evolution
in the variational description, are a finite set of variables which necessarily span a 
homogeneous space of the kinematical group.
The manifold they span is larger than the configuration space and, in addition
to the independent degrees of freedom, it also includes the derivatives of the 
degrees of freedom up to one order less than the highest order they have in the Lagrangian.
The Lagrangian for describing these systems will be thus dependent on these kinematical 
variables $x$ and their next order time derivative.
If the evolution is described in terms of some group invariant evolution parameter $\tau$, then,
when writting the Lagrangian not in terms of the independent degrees of freedom but as a function of 
the kinematical variables and their $\tau-$derivatives, $\dot{x}$,
it becomes a homogeneous function of first degree of the $\tau- $derivatives
of all kinematical variables. This feature will allow us to make a theoretical analysis without postulating 
any particular Lagrangian.

The formalism is completely general and can accomodate to any kinematical group we consider
as the spacetime symmetry group of the theory. 
But at the same time it is very restrictive, because once this 
kinematical group is fixed the kind of classical variables which define the initial and final states 
of an elementary particle in a variational approach, 
are restricted to belong to homogeneous spaces of the group. 
This kinematical group is the fundamental object of the formalism
and, therefore, we call the formalism {\it kinematical}, to stress this fact.

All elementary systems described within this formalism have the feature that, when quantized, 
their Hilbert space of pure states
carries a projective unitary irreducible representation of the kinematical group. It is through
Feynman's path integral approach that both formalisms complement each other.
For the Galilei and Poincar\'e groups, the most general homogeneous space is spanned by a set
of 10 variables, the same number and with the same geometrical interpretation 
as the group parameters, $(t,{\bi r},{\bi u},\balpha)$, 
interpreted as the time, position of the charge, velocity
of the charge and orientation, respectively. In the relativistic case we have three maximal, disjoint, homogeneous
spaces spanned by these variables, according to the value of the velocity $u<c$, $u=c$ and $u>c$. 
The quantization of the manifold with $u=c$,
produces Dirac's equation \cite{Rivasbook}. If $x\equiv(t,{\bi r},{\bi u},\balpha)$
are the kinematical variables, then the Lagrangian will also depend on the acceleration and on
the angular velocity. The dynamical equations for the point ${\bi r}$ will be, in general, of fourth
order.

A general spinning elementary particle is just a localized and orientable mechanical system. 
By localized we mean that to analyse its evolution in space we have just to describe the 
evolution of a single point ${\bi r}$,
where the charge is located and in terms of which the possible interactions are determined. 
This point  ${\bi r}$ also represents the centre-of-mass of the system for spinless particles, 
while for spinning ones must necessarily be a different
point than ${\bi q}$, the centre-of-mass, very well defined classically 
and where we can locate the mass of the particle. 
It is the motion of the charge around the centre-of-mass which gives rise to a 
classical interpretation of the {\it zitterbewegung} and also to the
dipole structure of the particle.
By orientable we mean that in addition to the description of the evolution of the point charge we also need to
describe the change of orientation of the system $\balpha$, by analysing the evolution of a local comoving and rotating 
frame attached to that point. An elementary spinning particle is thus described as we use to describe a rigid
body but with some differences: we have not to talk about size or shape and the point does not represent
the centre-of-mass but rather the centre-of-charge. It is allowed to satisfy a fourth order 
differential equation and, for a Dirac particle, it moves at the speed of light. 

%%%%%%%%%%%%%%%%%%%%%%%%%%%%%%%%%%
\section{A Dirac particle} 
\label{sec:Dirac}
%%%%%%%%%%%%%%%%%%%%%%%%%%%%%%%%%%

This model of elementary spinning particle was already quantized through Feynman's path integral
method \cite{quantum} and shown to satisfy Dirac's equation. 
Therefore it corresponds to a classical spinning model of a spin 1/2 object when quantized. 
The classical expression which gives rise to Dirac's equation is 
\[
H={\bi P}\cdot{\bi u}+\frac{1}{c^2}{\bi S}\cdot\left(\frac{d{\bi u}}{dt}\times{\bi u}\right),
\]
where the energy $H$ is expressed as the sum of two terms, ${\bi P}\cdot{\bi u}$, or translational energy
and the other, which depends on the spin of the system, or rotational energy. This part can never vanish for any observer,
while the first one is zero for the centre-of-mass observer. The spin comes from the dependence of the Lagrangian
of both, the acceleration $\dot{\bi u}$, and the angular velocity  ${\bomega}$, and if we define
\[
{\bi U}=\frac{\partial L}{\partial \dot{\bi u}},\quad {\bi W}=\frac{\partial L}{\partial\bomega},
\] 
it takes the form
\[
{\bi S}={\bi u}\times{\bi U}+{\bi W}={\bi Z}+{\bi W}.
\]
The first part ${\bi Z}={\bi u}\times{\bi U}$, or {\it zitterbewegung} part, 
is related to the separation between the centre-of-charge from the centre-of-mass and 
takes into account this relative orbital motion. It quantizes with integer values. 
The second part ${\bi W}$
is the rotational part of the body frame and quantizes with both integer and half-integer values.
The total angular momentum with respect to the origin of observer's frame is
\[
{\bi J}={\bi r}\times{\bi P}+{\bi S},
\]
so that the spin ${\bi S}$ is the angular momentum of the system with respect 
to the centre-of-charge ${\bi r}$,
and not with respect to the centre-of-mass ${\bi q}$. This is the reason why for a free particle 
it is not a conserved quantity,
but it satisfies the dynamical equation
 \begin{equation}
\frac{d{\bi S}}{dt}={\bi P}\times{\bi u}.
 \label{eq:dynspin}
 \end{equation}
This is exactly the same dynamical equation satisfied by Dirac's spin operator in the quantum case. 
If at point ${\bi r}$
there is defined some external force ${\bi F}$, the total angular momentum is no longer conserved and thus
\[
\frac{d{\bi J}}{dt}={\bi r}\times{\bi F}={\bi r}\times\frac{d{\bi P}}{dt}+{\bi u}\times{\bi P}+\frac{d{\bi S}}{dt}
\]
and because ${d{\bi P}}/{dt}={\bi F}$, Dirac's spin ${\bi S}$ also satisfies the dynamical equation (\ref{eq:dynspin})
for an interacting particle. This has to be taken into account when comparing the analysis of this spin 
with other approaches, for instance, with Bargmann-Michel-Telegdi spin observable \cite{BMT}, which clearly represents the
angular momentum with respect to the centre-of-mass of the system.

When expressed the spin and the centre-of-mass position 
in terms of the kinematical
variables and their derivates, they take, respectively, the form
\[
{\bi S}=\left(\frac{H-{\bi u}\cdot{\bi P}}{\left({d{\bi u}}/{dt}\right)^2}\right)\,\frac{d{\bi u}}{dt}\times{\bi u},\quad 
{\bi q}={\bi r}+\frac{c^2}{H}\left(\frac{H-{\bi u}\cdot{\bi P}}{(d{\bi u}/dt)^2}\right)\frac{d{\bi u}}{dt}
\]
Dirac's spin is always orthogonal to the osculator plane of the trajectory of the charge ${\bi r}$, 
in the direction opposite to the binormal for a positive energy particle, and the 
acceleration is pointing from ${\bi r}$ to the centre-of-mass, like in a helix.
It is shown that the dynamical equation of point ${\bi r}$ for the free particle and 
in the centre-of-mass frame is given by
 \begin{equation}
{\bi r}=\frac{1}{mc^2}{\bi S}\times{\bi u},
 \label{eq:dyq}
 \end{equation}
and where the spin vector ${\bi S}$ is constant in this frame, as depicted in Fig.~1.
The radius of the zitterbewegung motion is $R=S/mc$, and the angular velocity $\omega=mc^2/S$. 
When considered in the centre-of mass frame and all translational degrees of freedom are supressed, 
it is a system of three degrees of freedom; two are the 
$x$ and $y$ components of the position of the charge on the zitterbewegung plane and the third is the phase
of the rotation of the body frame. This phase is the same as the phase of the orbital motion and because
the velocity $u=c$ is constant, we are just left with a single and independent degree of freedom, for instance,
the $x$ coordinate. The Dirac particle, in the centre-of-mass frame, 
is a one-dimensional harmonic oscillator of frequency $\omega$. 

We can allow the system to have excited states, but the atomic principle 
suggests that an elementary particle cannot have excited states and that the only allowed 
state in the centre of mass frame, corresponds to the ground state. 
Its quantized ground energy $\hbar\omega/2$ is 
identified with the particle rest frame energy $mc^2$.
In this way, because $\omega=mc^2/S$, the classical spin $S$ takes the value $\hbar/2$, when quantized.
If the state of the particle, in the centre of mass frame, 
was some other excited state the value of the classical parameter $S$
will be different than $\hbar/2$, and thus contradictory with the condition 
that this system satisfies Dirac's equation. Therefore, the atomic hypothesis interpreted in the sense
that the system has no excited states produces the same result for the quantized spin 
than Feynman's quantization.

\cfig{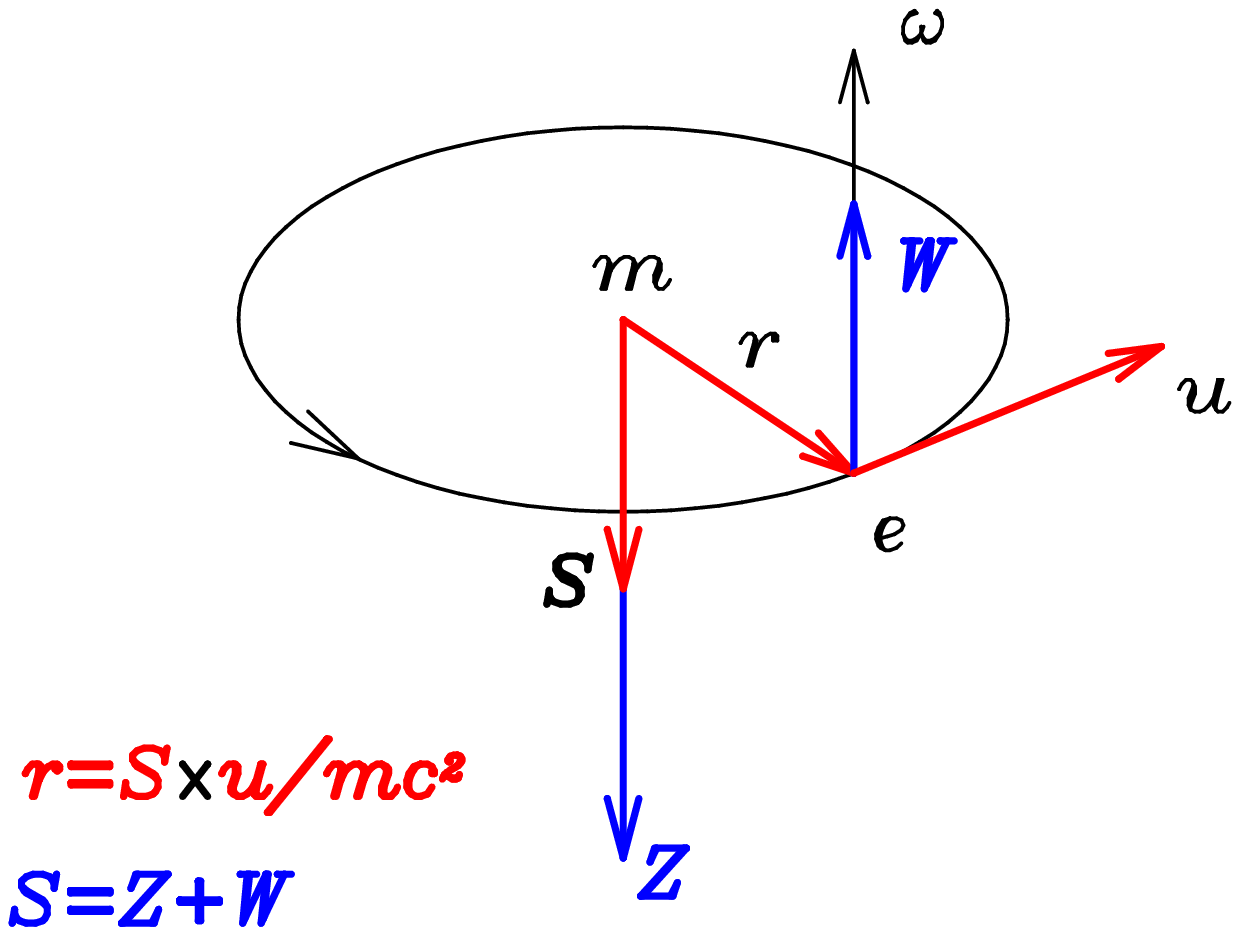}{Motion of the charge of the electron in the centre-of-mass frame. The magnetic moment
of the particle is produced by the motion of the charge. The total spin
${\bi S}$ is half the value of the zitterbewegung part ${\bi Z}$ when quantizing the sytem, so that when expressing the magnetic
moment in terms of the total spin we get a $g=2$ gyromagnetic ratio \cite{g2}. 
The body frame attached to the end of point ${\bi r}$, which could be the Frenet-Serret triad,
rotates with angular velocity $\bomega$, has not been depicted.}

When seen from an arbitrary observer (see Figure 2), the motion of the charge is a helix, 
so that according to (\ref{eq:dynspin}) Dirac's spin preccess around the direction of the 
conserved linear momentum ${\bi P}$. For a free particle, the centre-of-mass
spin
\[
{\bi S}_{CM}={\bi S}+({\bi r}-{\bi q})\times{\bi P},
\]
is a conserved quantity. The centre-of-mass velocity is ${\bi v}=d{\bi q}/dt$, and the linear momentum is written as usual
as ${\bi P}=\gamma(v)m{\bi v}$, so that the transversal motion of the charge
is at the velocity $\sqrt{c^2-v^2}$. A moving electron takes a time $\gamma(v)$ times longer
than for an electron at rest to complete a turn, as a result of the time dilation measurement.

\cfig{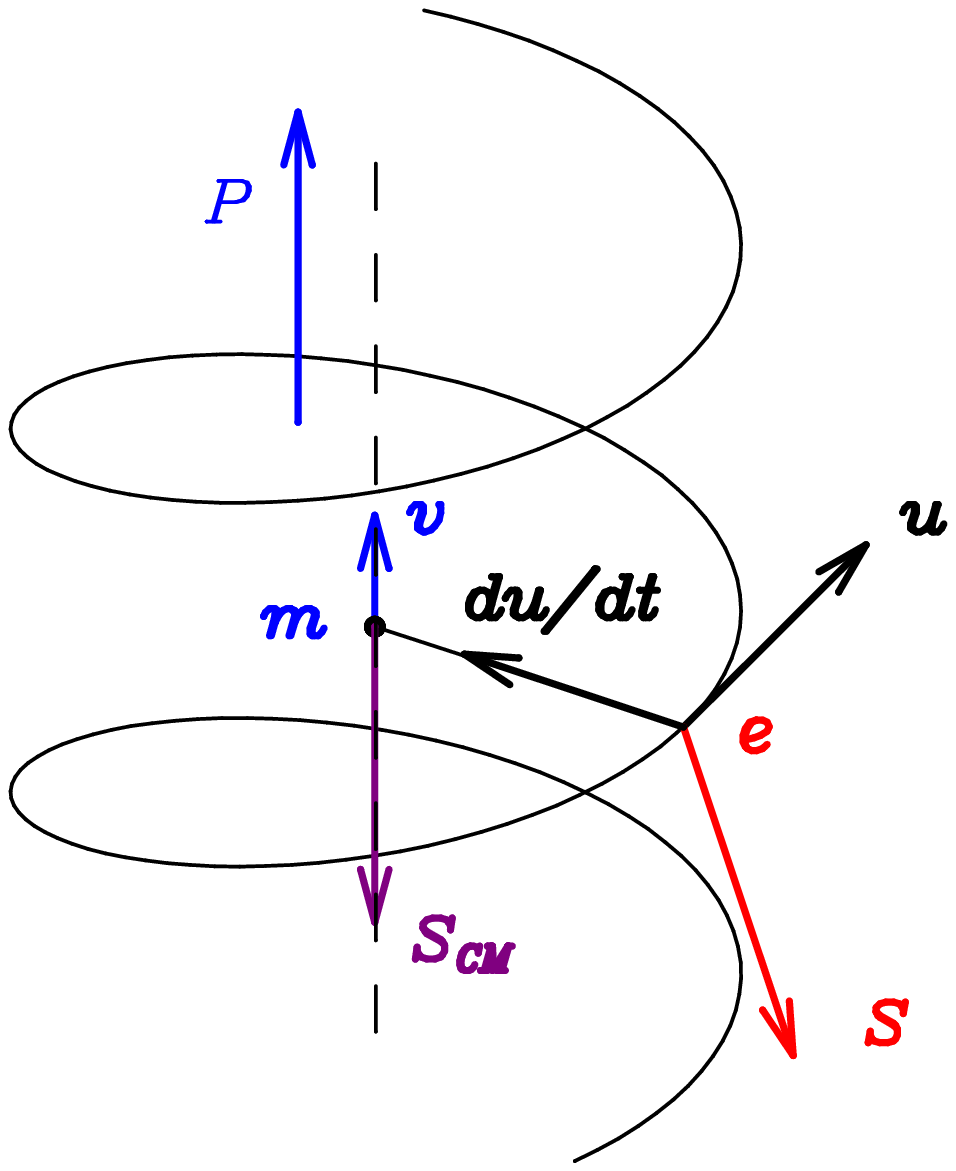}{Precession of Dirac's spin along the linear momentum ${\bi P}$. The tranversal
motion takes a time $\gamma(v)$ longer to complete a turn. The three vectors ${\bi u}$, $d{\bi u}/dt$
and $-{\bi S}$, properly normalized, form the Frenet-Serret triad of the motion of the charge. The spin with
respect to the centre-of-mass ${\bi S}_{CM}$, is a constant of the motion for the free particle.}

In a recent work \cite{Rivaspacetime} we have shown that the spacetime symmetry group
for a Dirac particle,
can be enlarged to include also spacetime dilations and local rotations of the body frame.
This group is ${\cal W}\otimes SO(3)_L$, 
where ${\cal W}$ is the Weyl group, i.e., the Poincar\'e group ${\cal P}$ enlarged with spacetime dilations and
$SO(3)_L$ is the group of local rotations of the body frame, which commutes with ${\cal W}$.
Because the Weyl group has no central extensions \cite{Boya}, 
the Lagrangian for a free Dirac particle is also invariant under this enlarged group.

If we consider this new group as the kinematical group of the theory, then the 
kinematical variables of a Dirac particle are reduced to time $t$, position of a point ${\bi r}$,
where the charge of the particle is located, its velocity ${\bi u}$ with the constraint
$u=c$, the orientation $\balpha$ which can be interpreted as the orientation of a local frame
with origin at point ${\bi r}$ and characterized by three parameters of a suitable parameterization
of the rotation group and, finally, a dimensionless scale $\beta$ of the internal motion of the charge
around the centre-of-mass. If the particle has spin $S\neq0$ and mass $m\neq0$, then a length scale factor
$R=S/mc$ and a time scale factor $T=S/mc^2$ can be defined, such that all kinematical variables for the
variational description can be taken dimensionless. It is this argument which justifies the enlargement of
the spacetime symmetry group, to include spacetime dilations which preserve the speed of light.

The Casimir operators of the enlarged group are the absolute value of the spin $S$, which is the Casimir
operator of the Weyl group ${\cal W}$, and the absolute value $I$ of the spin projection operator
on the body frame of the rotational part of the spin
\[
I_i={\bi e}_i\cdot{\bi W}
\]
which corresponds to the Casimir operator of the $SO(3)_L$ part. Here ${\bi e}_i$, $i=1,2,3$ represent
the three unit vectors of the local frame attached to the point ${\bi r}$.

A Dirac particle, with the enlarged group ${\cal W}\otimes SO(3)_L$ as its kinematical group, has as intrinsic properties
the spin $S$ and the spin projection $I$ which take both the eigenvalue $1/2$ when quantized \cite{Rivaspacetime}. 
By this reason, the four components of Dirac's spinor can be classified according to the $\pm1/2$ eigenvalues
of the $S_3$ and $I_3$ components of these spin operators.

%%%%%%%%%%%%%%%%%%%%%%%%%%%%%%%%%%
\section{Physical consequences} 
\label{sec:physics}
%%%%%%%%%%%%%%%%%%%%%%%%%%%%%%%%%%
Another important aspect of the atomic principle appears when we analyse the interaction between two Dirac particles
\cite{interac}. The general structure of the free Lagrangian is 
 \begin{equation}
L_0=T\dot{t}+{\bi R}\cdot\dot{\bi r}+{\bi U}\cdot\dot{\bi u}+{\bi W}\cdot\bomega+B\dot{\beta},
 \label{lag}
 \end{equation}
because the Lagrangian is a homogeneous function of first degree in terms of the derivativatives of the kinematical variables.
Here, $T={\partial L_0}/{\partial\dot{t}}$, ${\bi R}={\partial L_0}/{\partial\dot{\bi r}}$, ${\bi U}={\partial L_0}/{\partial\dot{\bi u}}$, 
${\bi W}={\partial L_0}/{\partial\bomega}$ and  $B={\partial L_0}/{\partial\dot{\beta}}$.
When we consider a compound system of two Dirac particles, the general Lagrangian will have the
form $L=L_1+L_2+L_I$, in terms of the free Lagrangians $L_1$ and $L_2$ for each particle and an interaction Lagrangian $L_I$. 
The free Lagrangian for each particle, will have the general 
form (\ref{lag}) in terms of the corresponding kinematical variables of each particle. The interaction Lagrangian
will be, in general, a homogeneous function of first degree in terms of the derivatives of all kinematical variables
of both particles. But if we assume the atomic principle, the internal structure of each particle cannot be modified.
This means that the spin $S$ and the spin projection on the body frame $I$ for each particle
have to be obtained only from the corresponding free Lagrangian. This forbids the dependence of the interaction
Lagrangian on the acceleration and angular velocity of the particles. A final invariance under spacetime dilations
to obtain a Lagrangian invariant under the new kinematical group ${\cal W}\otimes SO(3)_L$, gives rise to the
interaction Lagrangian
\begin{equation}
L_I=g\sqrt{\frac{c^2\dot{t}_1\dot{t}_2-\dot{\bi r}_1\cdot\dot{\bi r}_2}{({\bi r}_2-{\bi r}_1)^2-c^2(t_2-t_1)^2}},
 \label{eq:intL}
 \end{equation}
where $g$ is a coupling constant and the subindexes refer to the corresponding particles. This Lagrangian is clearly
invariant under the interchange $1\longleftrightarrow2$ of both particles.
When making a synchronous
desciption for any arbitrary observer, it becomes
\begin{equation}
L_I=g\sqrt{\frac{c^2-{\bi u}_1\cdot{\bi u}_2}{({\bi r}_2-{\bi r}_1)^2}}=g\frac{\sqrt{c^2-{\bi u}_1\cdot{\bi u}_2}}{r}
 \label{eq:li}
 \end{equation}
where $r=|{\bi r}_1-{\bi r}_2|$ is the instantaneous separation between the corresponding charges. 
When the spin of both particles is supressed, by taking in the low energy limit, the average 
values of the velocities of both charges will vanish, and the Lagrangian becomes
the instantaneous Coulomb Lagrangian between two point charges, thus suggesting that $gc=\pm e^2$.
It is with the use of this interaction Lagrangian that the formation of bound states will be analysed in
the next section.

In quantum electrodynamics the interaction Lagrangian between Dirac particles is obtained through
the local gauge invariance prescription for the Dirac field. 
This requirement predicts the existence of
a massless spin 1 field so that the interaction between Dirac particles is mediated
through the gauge photon field in the form $j^\mu A_\mu$, where the particle current
$j^\mu=e\bar{\psi}\gamma^\mu\psi$ is coupled to the electromagnetic field $A_\mu$.
In classical physics we have no means to describe, in a system of a finite number of degrees
of freedom, the possibility of changing the number of particles and how intermediate
particles could be created. We express the interaction only in terms of the classical variables associated
to each particle. The atomic principle has been used here to restrict, among the possible interaction
Lagrangians, those which do not modify the spin structure of any of the classical spinning particles.

To see if the classical Lagrangian (\ref{eq:intL}) describes something equivalent to the quantum mechanical
interaction Lagrangian of the Dirac field with the intermediate gauge field we have to proof that our classical 
Lagrangian can be rewritten, for instance,
in the form of a coupling of each particle current with the retarded classical electromagnetic potential
of the other, i.e., $j_1^\mu {A_{2}}_\mu+j_2^\mu {A_{1}}_\mu$ or something alike.
However, this Lagrangian describes an action at a distance interaction between particles
in the form of a coupling of the particles four-velocities $\dot{x}_1^\mu\dot{x}_{2\mu}$ and 
in terms of their spacetime separation $(x_1-x_2)^2$, but not
in terms of the retarded spacetime positions so that, 
if the above decomposition could be achieved, the Lagrangian could be interpreted as the predictivization of 
the retarded electromagnetic interaction between the two particles. 
This task is, probably, cumbersome and out of the scope of the present
section which tries to enhance the role of the atomic principle in restricting the allowed interactions.
It poses an interesting research subject for future work. But, nevertheless, 
the present Lagrangian contains as a limit, 
when the spins of the particles are supressed, the instantaneous Coulomb interaction 
between the two point charges, which is a nice and expected non-relativistic and spinless limit.

If we succeed in showing this feature it would mean some relationship between the quantum local gauge
invariance statement and the atomic principle because they lead, in the quantum and classical framework, 
respectively, to a similar interaction description. It is an interesting theoretical ansatz.

%%%%%%%%%%%%%%%%%%%%%%%%%%%%%%%%%%
\section{Predictions of the formalism} 
\label{sec:predict}
%%%%%%%%%%%%%%%%%%%%%%%%%%%%%%%%%% 
The formalism produces several predictions:
\begin{itemize}
\item{Chirality. Matter is lefthanded and antimatter righthanded.}
\item{Particles and antiparticles have the same relative orientation between the spin and magnetic moment.}
\item{A repulsive force betweeen charges does not forbid the formation of bound states, provided
the spins are parallel.}
\end{itemize}

\cfig{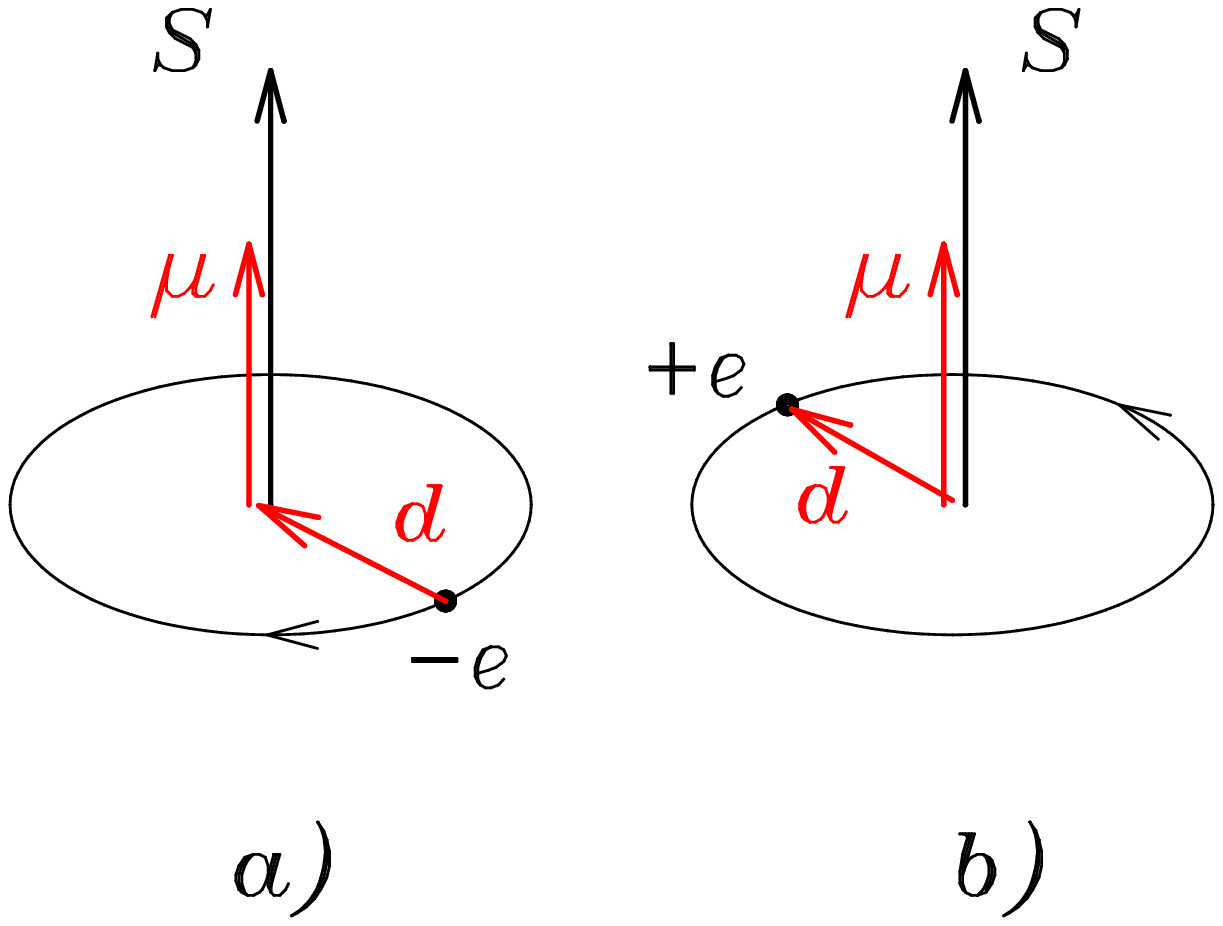}{(a) Motion of the charge, showing the electric and magnetic dipole with respect to the centre-of-mass, 
for the positive energy particle. (b) The $PCT$ transformed system corresponds to the antiparticle, with the same relative orientation
between spin and magnetic moment. Once the spin direction is fixed as a positive direction, 
the motion of the charge for the particle (a) is clockwise, 
which corresponds to a negative or left-handed motion. Antimatter (b), corresponds to a right-handed motion.
The instantaneous electric dipole with respect to the centre-of-mass is defined as ${\bi d}=\pm e({\bi r}-{\bi q})$.}

The conserved kinematical momentum, i.e., the constant of the motion associated to the
invariance of the Lagrangian under Lorentz boosts, takes the form for a Dirac free particle
\[
{\bi K}=H{\bi r}/c^2-{\bi P}t-{\bi S}\times{\bi u}/c^2.
\]
In the centre-of-mass frame ${\bi P}={\bi K}=0$ and $H=\pm mc^2$, and Dirac's spin ${\bi S}$ is
a constant vector, so that the motion of the charge is given by
\[
{\bi r}=\pm \frac{{\bi S}\times{\bi u}}{mc^2}, \quad (+\; {\rm particle}),
 \quad (-\; {\rm antiparticle}),
\]
and these two motions are depicted, respectively, in Figure 3(a) and 3(b), 
where we have chosen for the particle
the charge $-e$ and for the antiparticle $+e$. There is an arbitrariness in the selection
of the charge of the particle, but the motion of the positive energy solution is clockwise,
once the spin direction is fixed, while for the antiparticle we have a counterclockwise motion,
although both motions produce the same spin.
Then, the particle makes a negative trajectory in the zitterbewegung plane, 
thus representing a left-handed system.
Antimatter moves according to a right-handed system.

This produces that particle and antiparticle have the same magnetic dipole with the same relative 
orientation with respect to the spin. 
They also have an instantaneous electric dipole which rotates very fast around the spin direction, 
so that its time average is basically zero for low energy proccesses.
The electron, as an average, can thus be considered as a point charge at rest and some magnetic dipole, 
located both at the centre-of-mass. But in very close electron-electron interaction or in 
high energy processes, both electric and magnetic dipoles have to be taken into account for describing
the interaction, or, alternatively, the knowledge of the actual location of the corresponding charges.

As a matter of fact, the positronium (electron-positron bound sytem) has a ground state
of spin 0 and magnetic moment 0. This means that the spins of both electron and positron
are antiparallel to each other and the same thing happens to the corresponding magnetic moments. 
Therefore, for the
electron and positron there would exist the same relative orientation between the spin and 
magnetic moment.
The neutral pion $\pi^0$ is a linear combination of the quark-antiquark bound systems $u\bar{u}$, $d\bar{d}$
and sometimes the pair $s\bar{s}$ is also included. It is a system of 0 spin and 0 magnetic moment. Because
each of the above quarks have different masses and charges, and thus different magnetic moments, 
the possibility is that each quark-antiquark pair is a system of 0 spin and 0 magnetic moment, 
and, therefore each quark
and the corresponding antiquark must have the same relative orientation between 
the spin and magnetic moment.

This feature is opposite to what is usually assumed because for the electron it is taken that spin 
and magnetic moment are opposite to each other, while for the positron they are taken parallel. However,
in my opinion, there is no clear experimental evidence in the literature of this fact and, therefore,
experimentalists should check at least, for electrons and muons, 
whether they have the spin and magnetic moments
parallel or antiparallel. One possibility is to analize the motion of these particles in storage
rings. If, as predicted, they have the same relative orientation, then when injecting in the same direction,
$e^+$ and $e^-$ (or muons either), polarized in the up direction, 
their spins must preccess in the opposite direction, because
the magnetic field has to be reversed when we change from particles to antiparticles. 
The direction of preccession will show whether they are parallel or 
antiparallel. 

If we analyse the interaction of two Dirac particles we can use the mentioned 
interaction Lagrangian (\ref{eq:li}) which is invariant under the enlarged ${\cal W}\otimes SO(3)_L$ group. 
When the two particles are far apart,
the behaviour of the interaction becomes the instantaneous Coulomb interaction between the charges. 
In Figure 4 we represent the sccatering of two electrons with the spins parallel and where 
the trajectories of the corresponding centres-of-mass
are also depicted. In this example the two particles approach each other to a 
separation greater than Compton's wavelength.
\cfig{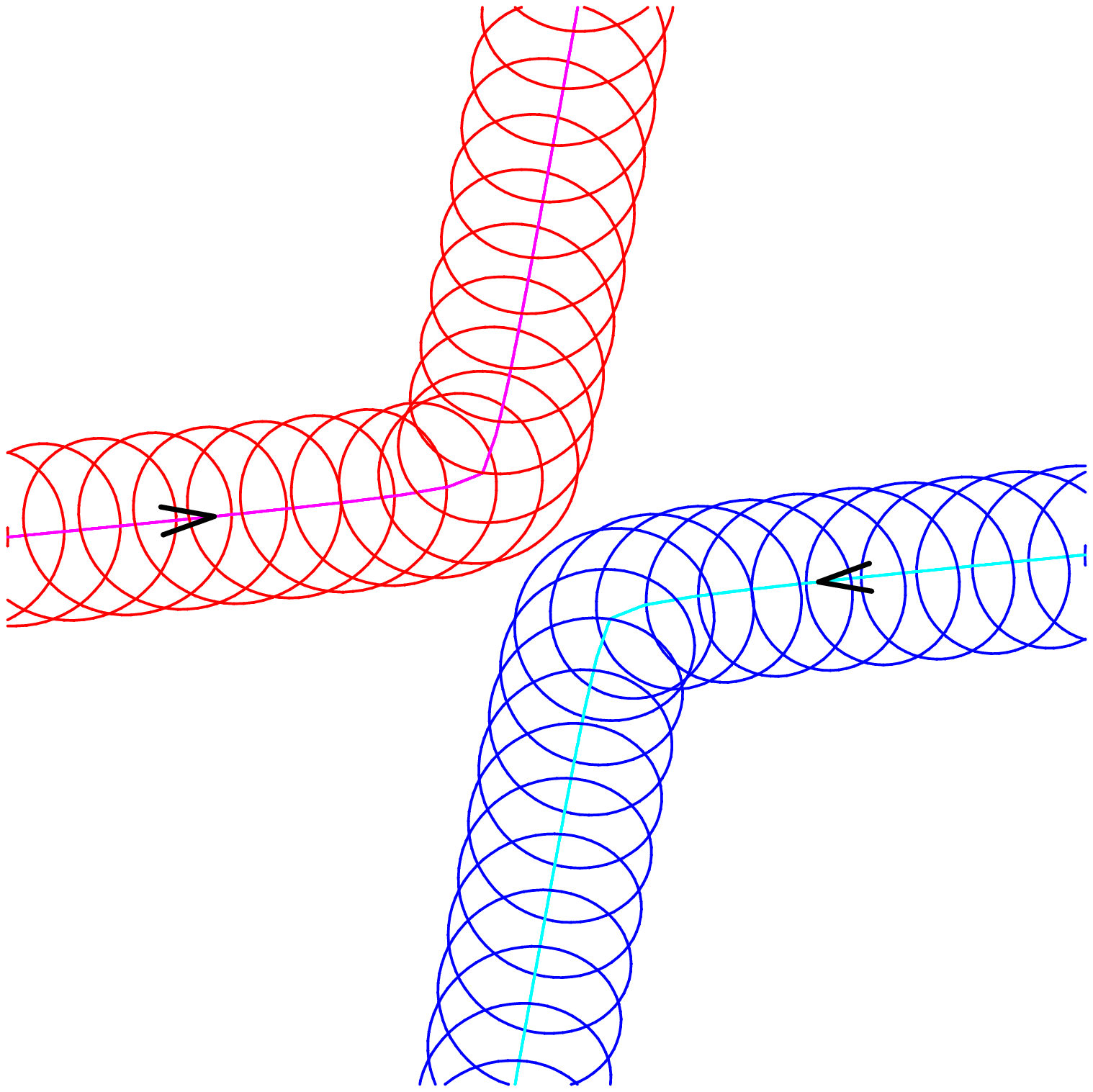}{Sccatering of two electrons with their spins parallel. The intermediate trajectories,
marked with an arrow, correspond to the evolution of the corresponding centres-of-mass.}

But if we locate very closely the two electrons, below Compton's wavelength, 
provided the phases of the charges
in the internal motion are opposite to each other, and the velocities below $0.01c$, 
we can obtain metastable bound motions
like the one depicted in Figure 5. The mass of this spin 1 bound system is greater 
than $2m_e$, because the potential
and kinetic energies are both positive. The solution of the corresponding 
quantum analysis, in particular
the possible quantization of the binding energies, is not yet done. The analysis of this 
bound motion has been done in \cite{interac}. 

\cfig{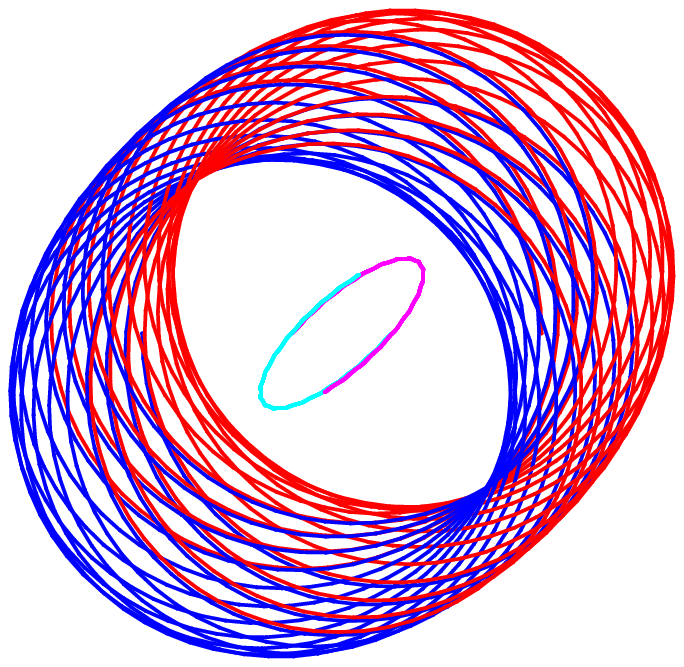}{Bound motion of two electrons with the spins parallel, for an initial velocity of the centre-of-mass
of each particle of $v<0.01c$. The phases of the charges have to be, basically, opposite to each other to produce
a metastable bound system, and the initial separation between the centres-of-mass is $0.2\times$Compton's wavelength.}

To justify how two particles of the same charge can attract each other, we have to solve a 
system of fourth order
differential equations for each particle or, alternatively, a system of second order 
differential equations once
the centers-of-mass of the particles are defined. For each centre-of-mass trajectory 
we need to know the external force
acting on the corresponding particle, but this force is defined at the corresponding 
centre-of-charge, and as we see in Figure 6
\cfig{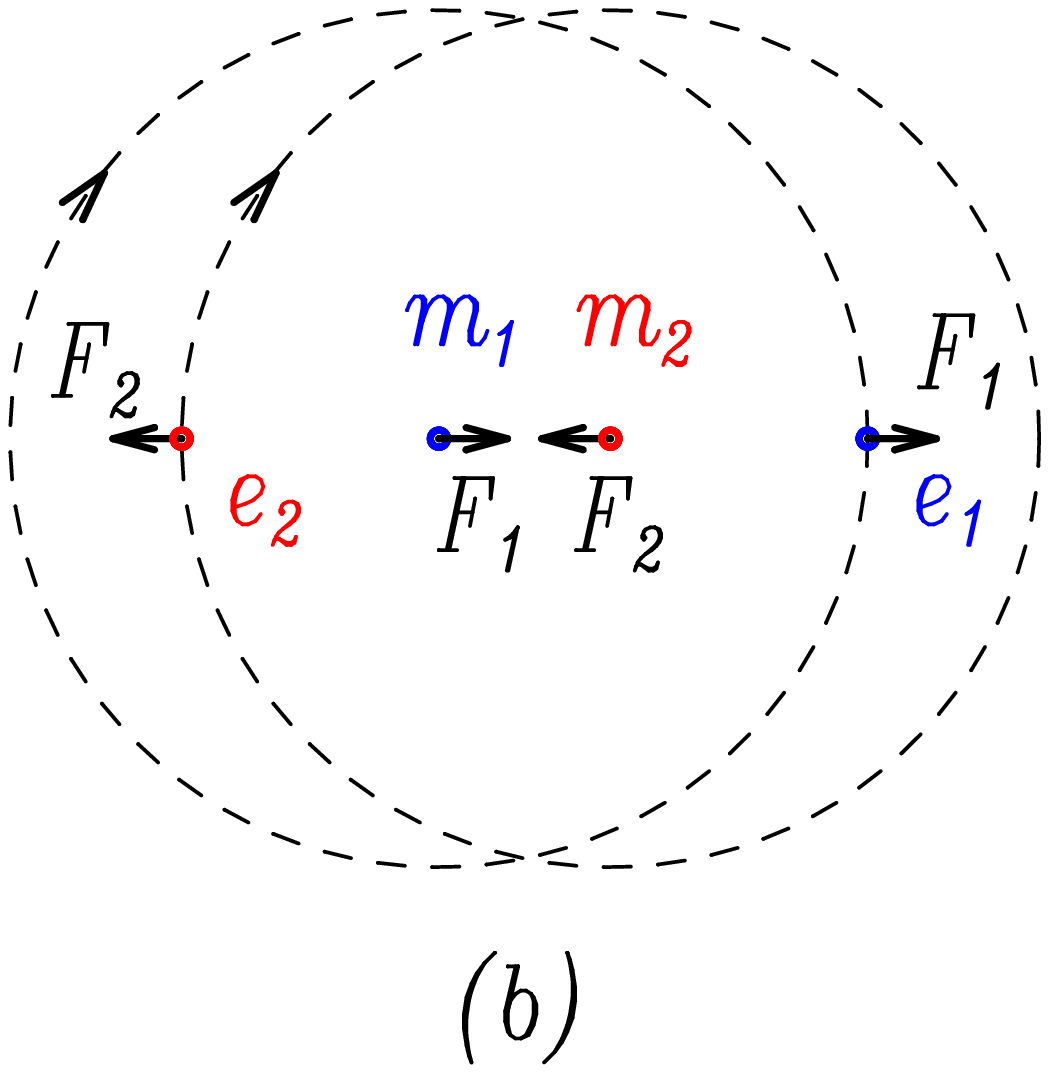}{Initial position of the charges of the two particles with the spins parallel and 
with a phase shift of $180^\circ$. The dotted lines represent the previsible 
evolution of each charge, in the same direction
for both particles, which implies that the spins are parallel. The repulsive force between charges
is also depicted at the corresponding centre-of-mass, 
thus producing an atractive force between the particles.}
a repulsive force between the charges implies an attractive force between the centres-of-mass provided the phases
of the charges are opposite to each other.

%%%%%%%%%%%%%%%%%%%%%%%%%%%%%%%%%%
\section{Summary and Conclusions}
%%%%%%%%%%%%%%%%%%%%%%%%%%%%%%%%%%

In a schematic form we list briefly some general features and conclusions about the kinematical formalism,
obtained by assuming the atomic hypothesis as a fundamental principle.

\begin{itemize}
\item{An elementary particle is a system without excited states. If it is not destroyed, its internal structure
can never be modified. All its possible states are kinematical modifications of any one of them.}
\item{The most general trajectory of the charge of a free elementary spinning particle is a helix at the speed of light.}
\item{The kinematical group supplies the symmetries and the variables for the variational description 
of an elementary particle, wich necessarily span a homogeneous space of the group.}
\item{These classical variables define the support manifold of the Hilbert space when quantizing the system.}
\item{The kinematical formalism is complete in the sense that the quantization of the models
produces all known one-particle wave-equations.}
\item{The spinning particles are localized and orientable systems.}
\item{The center-of-charge and center-of-mass are necessarily different points.}
\item{Elementary Dirac particles have a definite chirality. Matter is left-handed and antimatter right-handed.}
\item{The spin has a twofold structure: One part is related the orbital motion of the center-of-charge and the other
is related to the rotation of the particle.}
\item{This twofold structure produces a kinematical interpretation of the gyromagnetic ratio.}
\item{The magnetic moment is produced by the motion of the center-of-charge around the center-of-mass (zitterbewegung).}
\item{A particle and its corresponding antiparticle have the same relative orientation between the spin and magnetic moment.}
\item{The spacetime symmetry group of a Dirac particle is larger than the Poincar\'e group. It becomes,
at least, ${\cal W}\otimes SO(3)_L$.}
\item{It is the spin the only intrinsic property of a Dirac particle if considered under this kinematical group.}
\item{Two equal charged particles can form, from the classical point of view, bound systems provided
their spins are parallel and if their separation is below Compton's wavelength.}
\end{itemize}

\ack{I thank my colleague J M Aguirregabiria for the use of his excellent
Dynamics Solver program \cite{JMA} with which the numerical computations of the electron dynamics
have been done. This work has been partially supported by 
Universidad del Pa\'{\i}s Vasco/Euskal Herriko Unibertsitatea grant  9/UPV00172.310-14456/2002.}

%%%%%%%%%%%%%%%%%%%%%%%%%%%%%%%%%%%%%%%%%%%%

\end{document}